\documentstyle[prb,epsfig,twocolumn,aps]{revtex}
\def\etwa{\mbox{${}_{\sim}$}}

\newlength{\blib}

\begin{document}
\draft
\title{The optical response of Ba$_{1-x}$K$_x$BiO$_3$: 
Evidence for an unusual coupling mechanism of superconductivity?}
\author{H.\ J.\ Kaufmann}
\address{Interdisciplinary Research Centre in Superconductivity,
University of Cambridge, Madingley Road, Cambridge CB3 0HE, UK and
Department of Earth Sciences, University of Cambridge,
Downing Street, Cambridge CB2 3EQ, UK.}
\author{Oleg V Dolgov\cite{current}}
\address{P. N. Lebedev Physical Institute, 117924 Moscow, Russia.}
\author{E.\ K.\ H.\ Salje}
\address{Interdisciplinary Research Centre in Superconductivity,
University of Cambridge, Madingley Road, Cambridge CB3 0HE, UK and
Department of Earth Sciences, University of Cambridge,
Downing Street, Cambridge CB2 3EQ, UK.}
\maketitle
\begin{abstract}
\hspace*{1cm} We have analysed optical reflectivity data for Ba$_{1-x}$K$_x$BiO$_3$
in the far--infrared region using Migdal--Eliashberg theory and found it
inconsistent with standard electron--phonon coupling: 
Whereas the superconducting state data could be explained using moderate 
coupling, $\lambda=0.7$, the normal state properties indicate $\lambda\le 0.2$.
We have found that such behaviour could be understood using a simple model 
consisting of weak standard electron--phonon coupling plus weak coupling to an
unspecified high energy excitation near $0.4$ $e$V.
This model is found to be in general agreement with the reflectivity data,
except for the predicted superconducting gap size. The additional
high energy excitation
suggests that the dominant coupling mechanism in Ba$_{1-x}$K$_x$BiO$_3$ is 
not standard electron--phonon.
\end{abstract}
%\section{Introduction}
The high--${\rm T_c}$ oxide superconductor Ba$_{1-x}$K$_x$BiO$_3$ (BKBO)
has attracted much attention in recent years.
Its relatively high ${\rm T_c}$ ($\sim 30$ K), cubic structure in the
superconducting phase, and availability in thin film form has made
it an attractive candidate for microwave and tunnel junction devices.

From a theoretical point of view BKBO is interesting because it shows 
features of both the classical superconductors and the cuprates. 
Its parent compound BaBiO$_3$ has the perovskite structure. It is
a charge--density--wave insulator, similar to the undoped cuprates which 
are antiferromagnetic insulators.\cite{pei} BKBO shares with the
cuprates an asymmetric background in tunnelling conductance,\cite{zwei}
the overall behaviour of the reflectivity and
energy loss function,\cite{bozovic} and a very high ratio of ${\rm T_c}$
to the density of states at the Fermi level $N(0)$. The latter
is an extremely unfavourable situation for phonon--mediated
superconductivity.
On the other hand, BKBO has no layered structure, its parent compound is 
diamagnetic,\cite{cavas} and the charge carriers are  
electrons.\cite{fuenf,sechs} It has a well--defined gap feature in the optical 
\cite{sieben,puchkov} and tunnelling data \cite{neun,zehn,samueli} and 
shows a sizable isotope effect \cite{elf}.
These latter properties suggest that BKBO is a classical electron--phonon
s--wave superconductor. 

Within this standard mechanism moderate to strong coupling, $\lambda \sim 1$, 
is required to account for critical temperatures of ${\rm T_c}\sim 30$ K.
The observations of the gap size yield values for 
$2\Delta/k_{\rm B}{\rm T_c}$ between $3.5$ and $4.2$, allowing 
weak to strong electron--phonon coupling. 
Marsiglio {\it et al\/} [\onlinecite{marsiglio}] have analysed the imaginary part 
of the optical conductivity of BKBO from Ref.\ [\onlinecite{puchkov}]. They find 
that the electron--phonon coupling must be weak, $\lambda \approx 0.2$, to
explain the data.
Recent density functional calculations for BKBO yield an average 
value $\lambda=0.29$.\cite{savrasov} 
This could imply that standard electron--phonon coupling is not the dominant 
coupling mechanism in BKBO.

In this paper we have studied the far--infrared (FIR) reflectivity 
data for Ba$_{0.6}$K$_{0.4}$BiO$_3$ from Puchkov {\it et al\/} directly.
Analysing the data in the framework of standard Migdal--Eliashberg (ME)  
theory,\cite{migdal,eliashberg} 
we tried to find the simplest model that explains the observed data.
We found that the reflectivity in the superconducting state can be 
accounted for assuming moderate electron--phonon coupling, $\lambda = 0.7$.
However, in order to describe the normal state data, $\lambda \leq 0.2$ was required,
in agreement with Marsiglio {\it et al\/}.
A simple model that describes both normal and superconducting states,
was found by introducing an additional coupling to a high energy 
excitation near $0.4$ $e$V. 

\begin{figure}[ht]\begin{center}\leavevmode\epsfxsize=7cm
\epsfbox{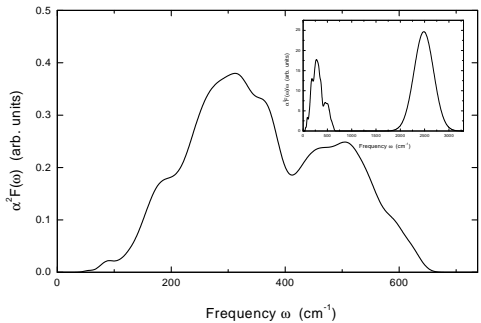}\end{center}
\caption{Eliashberg function $\alpha^2F$ used to describe the
electron--phonon interaction in BKBO. The inset shows 
$\alpha^2F(\omega)/\omega$ for a combined model consisting of
standard electron--phonon coupling plus a high energy excitation 
as described in the text.}
\label{fig1}
\end{figure}

The standard framework for describing the optical response of
high--${\rm T_c}$ superconductors is ME theory.
Disregarding vertex corrections, expressions for the optical
conductivity are well established in this strong coupling
extension of BCS theory and can be found
elsewhere.\cite{nam,Lee,dolgov}
They require the solution of the Eliashberg equations. 
Subsequently, the optical conductivity can be calculated in the 
local limit using the standard theory of the electromagnetic response function.
This approach can describe the overall behaviour of the optical response
of high--${\rm T_c}$ compounds 
%cuprates 
%YBa$_2$Cu$_3$O$_{7-\delta}$ and La$_{2-x}$Sr$_{x}$CuO$_4$
in the normal state.\cite{preprint}
To explain their optical response above and below ${\rm T_c}$ in detail
temperature dependent mid--infrared (MIR) bands should be included, 
as we have recently shown for La$_{2-x}$Sr$_{x}$CuO$_4$ (LSCO).\cite{holger}
The measured reflectivity from Ref.\ [\onlinecite{puchkov}] shows notable 
temperature dependence only in the FIR region. A simple model
can therefore consist of two components: a) a free carrier 
contribution treated within ME theory and b) a temperature
independent MIR band. The latter was mainly included for completeness 
here since we tried to explain the observed temperature dependence
in the FIR region. We modelled the MIR band by a simple Lorentzian
dielectric function.
%ME theory is no first principle theory but
%requires material parameters as an input.
To model the standard electron--phonon interaction we used an 
Eliashberg--function $\alpha^2F(\omega)$ derived from tunnelling 
data \cite{samueli}, which is shown in Fig.\ \ref{fig1}. 
Similar results were obtained by several groups of
investigators and there is good agreement with measured
phonon density of states for BKBO.\cite{huang} 
%and the overall shape of $\alpha^2F(\omega)$ in Fig.\ \ref{fig1} 
%can be taken as typical for BKBO.
Measurements of the Hall number \cite{lee} and the DC resistivity 
\cite{hellman} suggest a free carriers plasma frequency 
$\omega_{\rm P} \approx 15000$ cm$^{-1}$ and an impurity scattering rate 
$1/\tau_{\rm imp} \approx 150$ cm$^{-1}$.
The latter is close to $1/\tau_{\rm imp} \sim 180$ cm$^{-1}$ which
is the value derived from a simple Drude fit of the optical data. 
Here we used $1/\tau_{\rm imp} = 180$ cm$^{-1}$ and 
treated the plasma frequency as a fit parameter.
The electron--phonon coupling constant
\begin{equation}
\lambda=2\int\limits_{0}^{\infty}\frac{{\rm d\/}\omega}{\omega}\:
\alpha^2F(\omega)
\end{equation}
should be adjusted to yield the correct value for the superconducting
transition temperature ${\rm T_c}=28$ K.

\begin{figure}[ht]\begin{center}\leavevmode\epsfxsize=7cm
\epsfbox{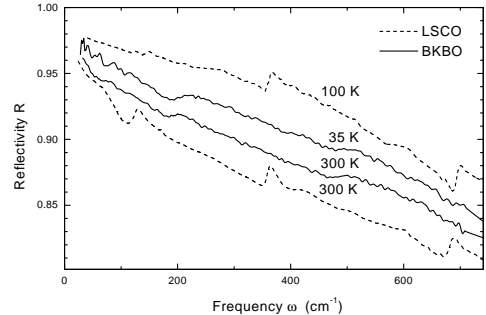}\end{center}
\caption{Comparison between measured FIR reflectivity of
LSCO and BKBO at different temperatures.}
\label{fig2}
\end{figure}

However, $\lambda$ is directly linked to the temperature dependence 
of the FIR spectra: Strong temperature dependence of the
FIR reflectivity indicates large $\lambda$ and vice versa.
It is important to note that this correlation is an intrinsic property
of ME theory which is not affected by the choice of $1/\tau_{\rm imp}$ and
$\omega_{\rm P}$.
Figure \ref{fig2} compares normal state reflectivity curves for
BKBO \cite{puchkov} and LSCO \cite{gao} at different temperatures. 
Transition temperature, plasma frequency and 
impurity scattering rate are comparable in these materials.
Although the overall behaviour of the reflectivity is qualitatively
similar for these two compounds, there is a quantitative difference in their
temperature dependence. The normal state optical response of
LSCO could be accounted for using $\lambda = 1.0$,\cite{holger} but 
the temperature dependence of the BKBO data is far weaker, suggesting
$\lambda \leq 0.2$.

\begin{figure}[ht]\begin{center}\leavevmode\epsfxsize=7cm
\epsfbox{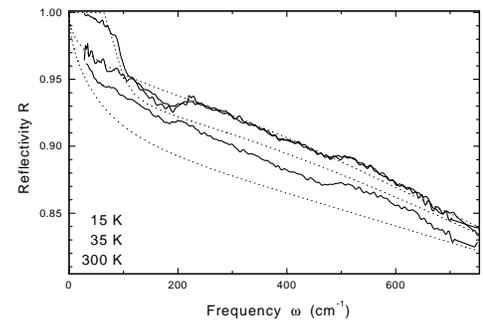}\end{center}
\caption{Measured (solid) FIR reflectivity of BKBO and 
calculated curves (dotted) using
$\lambda=0.7$ and $\mu^*=0$.}
\label{fig3}
\end{figure}

The superconducting state of BKBO, on the other hand, is well described
using moderate electron--phonon coupling, $\lambda = 0.7$. This value 
yields the correct ${\rm T_c}$ for the Eliashberg function shown in 
Fig.\ \ref{fig1}, assuming that the Coulomb pseudo--potential vanishes,
$\mu^*=0$, and 
it is in good agreement with far--infrared (FIR) transmission 
measurements at low temperatures.\cite{dunmore}
Figure \ref{fig3} shows the calculated FIR reflectivity for these
parameters, using a plasma frequency $\omega_{\rm P}=12240$ cm$^{-1}$ 
and $\epsilon_\infty=3.0$, and compares them to the measured curves.
% and a MIR band with $\omega_{\rm P_{MIR}}=6760$ cm$^{-1}$, 
%$\omega_{0}=2470$ cm$^{-1}$ , and $\Gamma=8710$ cm$^{-1}$
There is good agreement between model and experiment at 
$15$ K and $35$ K, but at room temperature the model underestimates
the measured reflectivity.

Therefore, the temperature dependence of the normal state data
suggest weak electron--phonon coupling, whereas a fit of the 
low--temperature data can be achieved assuming moderate coupling.
A possible explanation for such behaviour could be a strongly 
temperature dependent coupling constant $\alpha=\alpha({\rm T})$ 
in the Eliashberg function $\alpha^2F(\omega)$.
In an early Raman study on Ba$_{0.6}$K$_{0.4}$BiO$_3$, McCarty 
{\it et al\/} [\onlinecite{McCarty}] found a phonon peak at $325$ cm$^{-1}$
which developed a distinctive Fano line shape upon cooling,
which could support this picture.
Different effective coupling strength above and below ${\rm T_c}$
can arise if the electron--phonon interaction depends strongly on
momentum. In this case the normal state can be described by 
$\lambda=\lambda_0-\lambda_1$, where the $\lambda_i$ are Legendre components.
In the superconducting state the gap function $\Delta(\omega)$ 
and renormalisation factor $Z(\omega)$ are determined by $\lambda_0$,
whereas only $\lambda_1$ enters into the expression for vertex
corrections. In the simplest approach, neglecting vertex corrections, 
this notion results in different coupling constants above and below 
${\rm T_c}$. However, we found that this model can only account for
the experimental curves if one also allows for different plasma frequencies
and different MIR bands above and below ${\rm T_c}$.
Whereas the difference in $\omega_{\rm P}$ could arise from neglecting
vertex corrections below ${\rm T_c}$, it is difficult to see how this
could effect the MIR band. We therefore rule out 
strong momentum dependence of the electron--phonon interaction 
as an explanation of the observed behaviour, at least in the simplest
approximation. A full treatment, including vertex corrections, is beyond the 
scope of the present paper.

Another way to reconcile weak electron--phonon interaction with the
observed ${\rm T_c}$ would be the existence of an additional effective
electron--electron interaction. For the cuprates such an approach was
proposed very early on \cite{mars} but to date no conclusive evidence 
has been produced in its favour.
If the energy of the additional coupling mode lies in the region of phonon 
frequencies it cannot be distinguished from standard phonon modes. 
It would have to be strongly coupled to the electrons to account for the 
observed ${\rm T_c}$ and should be visible in tunnelling spectra.
Thus, if an additional contribution exists it must be at energies
well above the maximum phonon frequency $\omega_{\rm ph}\approx 700$
cm$^{-1}$. In the ME approach using only standard electron--phonon 
interactions such a high--energy mechanism would result in a negative
Coulomb pseudo--potential. For BKBO we found that the normal state 
reflectivity and ${\rm T_c}$ can indeed be 
reproduced assuming $\lambda=0.1$, $\mu^*=-0.074$, 
$\omega_{\rm P}=15080$ cm$^{-1}$, and $\epsilon_\infty=3.1$. These
parameters are in good agreement with the results of Ref.\ [\onlinecite{marsiglio}].
%The MIR band has the parameters $\omega_{\rm P_{MIR}}=11060$ cm$^{-1}$, 
%$\omega_{0}=1270$ cm$^{-1}$ , and $\Gamma=8210$ cm$^{-1}$.
Weak standard electron--phonon coupling is also in agreement with
the observed isotope effect in BKBO. For weak to moderate coupling strength one 
can use the phenomenological equation \cite{Carbotte}
\begin{equation}\label{alpha}
\alpha=\frac{1}{2}\,\left[\,
1-\frac{1.04(1+\lambda)(1+0.62\lambda)}{\left[
\lambda-\mu^*(1+0.62\lambda)\right]^2}\,\mu^{*\, 2}\right]
\end{equation}
which is based on McMillan's approximate formula for ${\rm T_c}$.
Loong {\it et al} report $\alpha=0.42\pm 0.05$. 
Moderate coupling, $\lambda=0.7$, would then require $\mu^* = 0.13$, 
which yields a critical temperature of only $15$ K,
whereas weak coupling, $\lambda=0.1$, suggests $\mu^*=0.26$ or a 
substantial negative Coulomb pseudo--potential, $\mu^*=-0.059$, 
indicating again the presence of a high--energy contribution.
For $\lambda=0.1$ and $\mu^*=-0.074$, Eq.\ (\ref{alpha}) predicts
$\alpha=0.40$, which is well within the error margins of Ref.\ [\onlinecite{elf}].

\begin{figure}[ht]\begin{center}\leavevmode\epsfxsize=7cm
\epsfbox{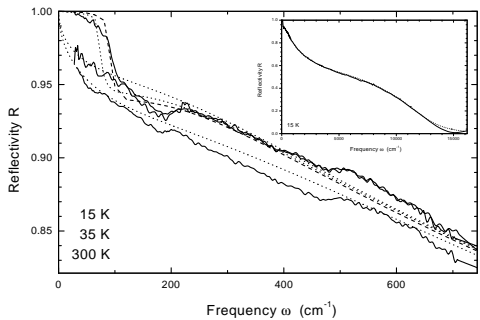}\end{center}
\caption{Measured (solid) and calculated (dotted) FIR reflectivity of BKBO,
assuming weak electron--phonon coupling plus an additional peak in 
$\alpha^2F(\omega)$ as described in the text.
The dashed line is the calculated curve at $15$ K for a stronger
high energy peak (${\rm T_c}=32$ K). The inset shows measured and 
calculated curves at $15$ K on a wider frequency scale.}
\label{fig4}
\end{figure}

\begin{figure}[ht]\begin{center}\leavevmode\epsfxsize=7cm
\epsfbox{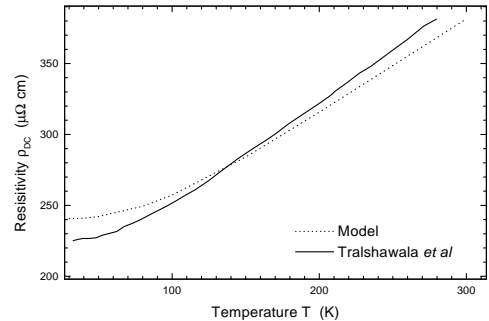}\end{center}
\caption{Measured and calculated DC resistivity for BKBO,
assuming weak electron--phonon coupling plus
an additional peak in $\alpha^2F(\omega)$ as described in the text.}
\label{fig5}
\end{figure}

We have studied a model consisting of two distinct contributions in
the Eliashberg function $\alpha^2F(\omega)$: The standard phonon
spectrum shown in Fig.\ \ref{fig1} plus a high energy excitation
centred at frequency $\Omega_{\rm e}>\omega_{\rm ph}$, modelled by
a Gaussian peak of width $400$ cm$^{-1}$. The electron--phonon coupling
was fixed to
\begin{equation}
\lambda_{\rm ph}=2\int\limits_{0}^{\omega_{\rm ph}}
\frac{{\rm d\/}\omega}{\omega}
\alpha^2F(\omega)=0.1
\end{equation}
and for simplicity we chose $\mu^*=0$. 
The respective coupling strengths of the standard phonon spectrum and
the high energy contribution can be observed in the inset of Fig.\
\ref{fig1} where $\alpha^2F(\omega)/\omega$ is plotted for a typical
case.
To derive ${\rm T_c}$ from
the calculated optical conductivity in this model, we calculated the
London penetration depth $\lambda_{\rm L}$, 
\begin{equation}
\lambda_{\rm L}^{-2}({\rm T})=\lim\limits_{\omega\rightarrow 0}
4\pi\omega\,{\rm Im}\, \sigma(\omega,{\rm T}),
\end{equation}
and extrapolated $\lambda_{\rm L}^{-2}({\rm T})$ for
${\rm T}\rightarrow {\rm T_c}$. We found that the normal state
reflectivity is unaffected by the existence of the high energy mode
as long as the total coupling is weak, $\lambda<0.4$. This puts 
a lower limit to $\Omega_{\rm e}$, $\Omega_{\rm e}>1500$ cm$^{-1}$.
Keeping ${\rm T_c}$ fixed, we varied $\Omega_{\rm e}$ between $2000$ 
cm$^{-1}$ and $7000$ cm$^{-1}$.
The superconducting state was little influenced by the position
of the high energy peak. The calculated gap size was unchanged
whereas the filling of the gap at ${\rm T} < {\rm T_c}$ decreased slightly
with increasing $\Omega_{\rm e}$. For $\Omega_{\rm e}> 7000$ cm$^{-1}$
the gap size started to decrease. There is therefore some ambiguity
in the position of the high energy excitation. For simplicity we
tried to keep $\Omega_{\rm e}$ small. Figure \ref{fig4} compares the calculated 
reflectivity for $\Omega_{\rm e}=2900$ cm$^{-1}$ to the experimental curves. 
The other parameters were $\lambda=0.34$, $\omega_{\rm P}=14530$ cm$^{-1}$, 
and $\epsilon_\infty=1.43$.
%$\omega_{\rm P_{MIR}}=6260$ cm$^{-1}$,$\omega_{0}=770$ cm$^{-1}$, and 
%$\Gamma=5100$ cm$^{-1}$
There is fair agreement between model and experiment in the normal state 
whereas in the superconducting state the calculated gap size
is somewhat too small. We found that for all $\Omega_{\rm e}$ the gap size 
is closer to the BCS limit, $2\Delta/k_{\rm B}{\rm T_c}=3.53$
than to the measured value, $2\Delta/k_{\rm B}{\rm T_c}=4.2\pm 0.5$.
This discrepancy is consistent with the fact that the overall coupling 
in our approach was still weak, $\lambda<0.4$. Even including a high energy 
excitation in the model requires $\lambda\ge 0.7$ to reproduce the 
measured gap size. 
The superconducting gaps reported in the literature vary from $3.5$ to $4.2$.
A possible explanation for the large gap observed in the reflectivity
could be found in the notorious inhomogeneity of BKBO in terms of potassium 
content and structural defects at a microscopic scale. i.\,e., the
${\rm T_c}$ probed locally in the optical experiment could be somewhat 
higher than the bulk value.
Also shown in Fig.\ \ref{fig4} is the calculated reflectivity at $15$ K 
assuming stronger coupling, $\lambda=0.35$, which corresponds to
${\rm T_c}=32$ K. Here, the observed gap size is accounted for.
We note that due to the presence of the high energy contribution in $\alpha^2F$
only small changes in $\lambda$ are required for notable changes in
${\rm T_c}$. 
%Experimentally the superconducting transition temperature was 
%obtained from magnetisation measurements \cite{puchkov} which showed
%an onset of superconductivity at $\sim 30.8$ K and a transition width
%of $\Delta{\rm T_c} \approx 3$ K. 
Generally, we found that in the superconducting state this model 
behaves like a standard weak coupling electron--phonon superconductor,
albeit with a high ${\rm T_c}$. 
The critical temperature agrees remarkably well 
with the prediction of the phenomenological McMillan formula,\cite{Carbotte}
\begin{equation}
{\rm T_c} = \frac{\omega_{\rm ln}}{1.2}\,{\rm exp}\left(
-\frac{1.04(1+\lambda)}{\lambda-\mu^*(1+0.62\lambda)}\right),
\end{equation}
where
\begin{equation}
\omega_{\rm ln}= {\rm exp}\left(\frac{2}{\lambda}\int\limits_{0}^{\infty}
\frac{{\rm d\/}\omega}{\omega}\:\alpha^2F(\omega)\,{\rm ln}\omega
\right)
\end{equation}
is a characteristic frequency in $\alpha^2F(\omega)$.
The penetration depth 
$\lambda_{\rm L}^2(0)/\lambda_{\rm L}^2({\rm T})$ follows the
phenomenological $1-({\rm T}/{\rm T_c})^4$ behaviour, and for $\lambda=0.34$
we found $\lambda_{\rm L}(0)=3570$ \AA, in good agreement with literature
values of $3300\pm 200$ \AA.\cite{zwoelf,pam} The calculated dielectric
loss function $-{\rm Im}\,\epsilon^{-1}(\omega)$ was proportional to
$\omega^2$ at low frequencies and had a (first) peak at
$1.54$ $e$V, in agreement with the results of Ref.\ [\onlinecite{bozovic}].
In Fig \ref{fig5} the calculated DC resistivity ($\lambda=0.34$) is compared
to the results from Ref.\ [\onlinecite{zasadzinski}] for a BKBO sample with gap 
value $\Delta_0=3.8$ m$e$V. Both curves show the same overall behaviour, 
indicating that the model discussed here is in general agreement with 
independent measurements.

Thus, the optical reflectivity of BKBO measured in Ref.\ 
[\onlinecite{puchkov}] cannot be explained by standard electron--phonon
coupling ME theory but suggests that the dominant mechanism 
is coupling to a high energy mode. It would be very interesting
to perform more FIR measurements on BKBO to confirm these
findings and to analyse them in more detail.  
The nature of the high energy excitation in our model is so far unclear.
A magnetic mechanism, widely favoured in the cuprates,
must be ruled out for BKBO since there are no local moments on any
of the ions involved. One possible candidate might be structural
fluctuations: superconductivity occurs in BKBO just above the
metal--insulator (MI) transition ($x=0.37$) which practically coincides 
with the orthorhombic (Ibmm) to cubic (Pm\=3m) phase transition. At low
temperatures, structural fluctuations could prevail in the cubic
phase and couple to the electrons. If these fluctuations involve
large clusters of ions they could have very high excitation energies.
This picture would explain why the maximum value of
${\rm T_c}$ is observed right after the MI transition and subsequently
decreases upon doping and it conforms with the discrepancy between the 
measured and calculated gap size.

In summary, we have studied the optical reflectivity of BKBO from
Ref.\ [\onlinecite{puchkov}] in the ME approach. We found that whereas
the low--temperature data suggested moderate electron--phonon coupling,
$\lambda=0.7$, the normal state data cannot be explained unless 
$\lambda\le 0.2$.
To account for such behaviour, we used a model consisting of 
weak standard electron--phonon coupling plus weak coupling to a high energy
excitation near $0.4$ $e$V. We found that this model behaves like a 
standard weak coupling electron--phonon superconductor, 
albeit with unusually high critical temperature. In contrast to
the measured reflectivity, the predicted superconducting gap size
is BCS--like. The model is in general agreement with independent 
measurements of the isotope effect, penetration depth, 
dielectric loss function, and DC resistivity in BKBO. 

%\section*{Acknowledgement}
The authors would like to thank A.\ V.\ Puchkov for kindly
providing his reflectivity data. They are also grateful to 
S.\ Shulga for providing his program for these calculations.

%\end{tighten}
\end{document}